\documentstyle[12pt,a4]{article}
\input{epsf.tex}
\begin{document}
\begin{center}
{\large {\bf Main parameters of cascade $\gamma$-decay of the $^{118}$Sn
 compound nucleus}}
\end{center}
\begin{center}{\bf J. Honz\'atko$^a$,
V. A. Khitrov$^b$, C. Panteleev$^b$, A. M. Sukhovoj$^b$}\\
{\sl $^a$ Nuclear Physics Institute, CZ-25068 \v{R}e\v{z} near Prague, Czech Republic}\\
{\sl $^b$Frank Laboratory of Neutron Physics, Joint Institute for Nuclear
Research \\ 141980 Dubna, Russia}
\end{center}

Main properties of the excited states of $^{118}Sn$ manifesting themselves
in cascade $\gamma$-decay of its compound state were studied. As in other
heavier nuclei studied earlier, qualitative interpretation of the totality
of the observed properties of this nucleus is impossible without accounting
for coexistence and interaction of quasi-particle and collective nuclear
excitations and their considerable influence on the main parameters of the
process under study.


\section{Introduction}

The study of the two-step $\gamma$-cascades following thermal neutron capture
in more than 50 nuclei of different types from $^{40}K$ to $^{200}Hg$ in
Dubna, Riga and \v{R}e\v{z} allowed us to obtain unique information on
properties of these nuclei in energy diapason from the ground state and,
practically, to the neutron binding energy $B_n$. Analysis of this information
provided the the following conclusions:

1.~Experimental intensities of the cascades to the groups of
low-lying levels $E_f$ cannot be reproduced in the calculation
with the precision achieved in the experiment if one uses
conventional model ideas of a nucleus. This cannot be done not
only within the simplest models like ``noon-interacting Fermi-gas"
but also in the framework of modern enough generalized model of
superfluid nucleus [1].

2.~Reliable enough energy dependence of level density and
radiative strength functions of dipole $\gamma$-transitions can be
extracted only from the experimentally obtained dependencies of
the cascade intensities on the energy of their primary transition
$E_1$ (or energy of their intermediate level $E_i=B_n-E_1$).

Practical possibility to solve this problem is determined by the
following factors:

(a)~Cascade intensity distribution (Fig.~1) is extracted [2] from
the mass of $\gamma -\gamma$ coincidences by means of the sum
coincidence method. It
 contains some number of pairs of narrow enough [3] full-energy peaks
 corresponding to intense two-step cascades and a ``noise line" with zero
 mean value formed by a number of low-intense cascades. The local specific
 deviations of the normal distribution of events in ``noise line" can be
 partially or completely rejected numerically [4].

 (b)~main part (more than 95-99\%) of intensity of cascades with  $E_1>0.5B_n$
 can be extracted from these spectra in form of energetically resolved pairs
 of peaks, quanta ordering in which is determined [5] with high reliability
 using the maximum likelihood method. But this requires modern enough [6]
  spectrometer for registration of coincidences.

Using only these data we determined (Fig. 2) most probable energy dependence
of all two-step cascades terminating at the ground and first excited states
of  $^{118}Sn$. This permits one to make according to method [7] quite
unambiguous (in limits of existing notions and possibilities to study this
process) conclusions about both density of the states excited at the thermal
 neutron capture and  reduced probability of their population by
 $\gamma$-quanta.

\section{Estimation of probable density $\rho$ of excited levels and
radiative strength functions $k$ of cascade transitions}

Radiative strength function
$k=\Gamma_{\lambda i}/(E_{\gamma}^3\times A^{2/3}\times D_{\lambda})$
(here $\Gamma_{\lambda i}$ is the partial width of $\gamma$-transition with the
 energy $E_{\gamma}$, $A$ is the nucleus mass and $D$ is the spacing between
 decaying levels $\lambda$) and level density $\rho$ determine the total
 radiative width of the compound state $\Gamma_{\lambda}$ and cascade
 intensity $I_{\gamma\gamma}$ [8] obtained in the following way:
\begin{equation}
\Gamma_{\lambda} =<\Gamma_{\lambda i}>\times m_{\lambda i}
\end{equation}

\begin{equation}
I_{\gamma\gamma}=\sum_{\lambda ,f}\sum_{i}\frac{\Gamma_{\lambda i}}
{\Gamma_{\lambda}}\frac{\Gamma_{if}}{\Gamma_i}=\sum_{\lambda ,f}
\frac{\Gamma_{\lambda i}}{<\Gamma_{\lambda i}> m_{\lambda i}}
n_{\lambda i}\frac{\Gamma_{if}}{<\Gamma_{if}> m_{if}}.
\end{equation}
Here the values of the total and partial gamma-widths are set for the
compound state $\lambda$ and cascade intermediate level $i$, respectively;
$m$ is the total number of the excited levels, and $n$ is the number of
excited levels in the energy interval $\Delta E$ of averaging of cascade
intensity.

These equations do not allow one to determine $k$ and $\rho$
unambiguously and independently. Some deviation of, for example,
$\rho$ from a real value is inevitably compensated by deviation of
strength functions of the corresponding magnitude and sign. As it
was shown in [7], however, possible value of these deviations is
small enough. Nevertheless, the results obtained in analysis [7]
can be used for the verification of nuclear models and, if
necessary, for the determination of the direction of the further
development of these models. The main argument in favour of this
statement is relatively week dependence of the final results on
the initial values of $k$ and $\rho$ (even if they are absolutely
unreal) in the iterative process [7]. The most serious supposition
of the described in [7] method for determination of $k$ and $\rho$
is the equality in energy dependence of radiative strength
functions for the primary and secondary transitions. If it is not
true, then the obtained values of $k$ and $\rho$ can have unknown
and, probably, significant systematic error. In dependence on the
sign of this error, discrepancy between the model and experimental
values of $k$ and $\rho$ can decrease or even increase.

Although calculation of the total $\gamma$-ray spectra with the use of level
density [7] and different (but, in principle, possible)
radiative strength functions showed [9] that
the main properties of level density are extracted with the high confidence
and probable error decreases discrepancy between the experimental and model
[10,11] values of $k$ and $\rho$, but independent complementary test of
application of the method [7] for each concrete nucleus is required.

\newpage

Some grounds to consider the $k$ and $\rho$ values, shown in
figures~3 and 4, as reflecting the most common regularities of
their energy dependences can be get from the comparison of the
experimental and calculated within different ideas of intensities
of cascades terminating at higher-lying levels (see Table). Taking
into account that the $k$ and $\rho$ values shown in figures~3 and
4 allow simultaneous reproduction of:

(a)~the total radiative width $\Gamma_\lambda$ of the compound state;

(b)~the energy dependence of the intensity $I_{\gamma\gamma}$
of cascades to the ground and
first excited states of $^{118}$Sn; and

(c)~the total intensity of the cascades to different final levels with the
energy up to 2.8 MeV\\
one can consider the strength functions and level density obtained
for $^{118}$Sn according to method  [7] as the most probable.
These data, of course, contain some systematic errors owing to
errors in determination of $I_{\gamma\gamma}$ and incompleteness
of the obtained information on the intensity of the two-step and
larger multiplicity cascades.

The results of the analysis are compared with predictions of the level
density models [11,12] and models of radiative widths [10,13].
In the case of radiative strength functions, a comparison is performed in
the following manner: the $k(E1)$ values calculated according to the
models [13] and [10] (upper and lower curves in Fig.~3, respectively) are
summed with $k(M1)=const$ which is normalized so that
the ratio $\Gamma(M1)/\Gamma(E1)$ would be approximately equal to the
experimental data at $E_{\gamma}\simeq B_n$.

A comparison of the results of the analysis with predictions of
the models [10-13] (often used for this aim) shows that:
\begin{enumerate}

\item the energy dependence of $k(E1)+k(M1)$ for $^{118}Sn$ differs
strongly from predictions of the models [10,13] as in the case of even-even
compound nuclei from the region of the $4s$-resonance of the neutron
strength function;

\item the probable level density conforms to the picture obtained in
previous experiments [11,12]: up to  the excitation energy $\simeq 3.5$ MeV,
our data do not contradict the exponential extrapolation of
$\rho (E_{ex})$ predicted by the Fermi-gas back-shift model [11].
 Level density  for $E_{ex}>3$ MeV is considerably less than that predicted
 by this model. Above
$E_{ex}\approx 5$ MeV the level density, most probably, better corresponds
to the predictions of the generalized model of the superfluid nucleus in
its simplest form [12]. The values of $\rho^{mod}$ predicted by the last
model decrease parameter $\chi^2=((\rho^{mod}-\rho^{best})/\delta \rho^{best})^2$
by a factor of about 3 as compared with the predictions of model [11] for
 the interval $E_{ex} \geq 4$ MeV.
\end{enumerate}

A very quick exponential increase in the level density above
$\simeq 5$ MeV says [12] about the probable dominant influence of
the inner, many-quasi-particle type of excitations of these
states.

\section{Discussion}
This conclusion is true only to a precision determined by existing
(and included in analysis [7]) notions of properties of the
excited states  and dynamics of development of the cascade
$\gamma$-decay process of neutron resonance.

The main of these notions consist in the following:

(a)~the branching coefficients at depopulation of any level $i$ do
not depend on mode of its population;

(b)~all levels from a given excitation energy interval follow the
sole statistic distribution. I.e., the mean reduced probability of
their population by primary $E1$ and $M1$ transitions is equal for
any level in the spin window determined by the selection rule and
does nod depend on the structure of wave functions of neutron
resonance and  intermediate cascade level $E_i$.

Therefore, according to the theorem on the average any sum of the
widths is represented in calculation of  $I_{\gamma\gamma}$ by
product of their number by the mean partial width (determined
through corresponding wave function).

 (c)~energy dependence of $k$ (but not its absolute value) is
 equal for the primary and secondary transitions of cascades.

 It is not known how the notions (a) and (b) of the $\gamma$-decay
 process of heavy nucleus are close to reality. This should be
 found from the experiment. Analysis of the excitation spectra of
 intermediate levels of the most intense cascades testifies to
 possibility of their equidistancy. Besides, it showes that the
 population of levels in $^{118}Sn$ above $\simeq 3$ MeV cannot be
 described in model calculation. These allow an assumtion about
 possible violation of, for example, notion (c). If interpretation
 of the results [14] as an existance of ``vibrational bands" built
 on the states with complicated structure corresponds to reality,
 then the enhanced $\gamma$-transitions with the energy of some
 hundreds keV and higher inside these bands and
 between them are quite possible, as well. Discrepancy of the
 results obtained according [7] and shown in Fig.~4 to the
 estimation [15] of the number of omitted levels admits as one of
 possible explanations the violation of notion (c).

\section{Conclusion}

The results of a comparison between the experimental and
calculated cascade intensities in this nucleus (and like in the
nuclei studied earlier) indicate a necessity to modify model
notions of the properties of the excited states of the heavy
nuclei. In the framework of the modern theoretical notions the
qualitative explanation of the obtained discrepancy between the
experiment and calculation can be removed only within the more
detailed accounting  by the nuclear models [1,12] for co-existence and
interaction of fermion and boson excitations of nuclear matter.
Otherwise, an achievement of complete correspondence between the
observed and calculated parameters of nuclear reactions, for
instance, neutron-induced reaction is impossible. This concerns,
partially, the total radiative widths of neutron resonances,
$\gamma$-spectrum, and cross-sections of neutron interactions.

This work was supported
by GACR under contract No. 202/03/0891 and
 by RFBR Grant No. 99-02-17863.
\\\\
The authors are thankful to Ivo Tomandl for his contribution in
the experiment carried out in \v{R}e\v{z} and for useful
discussion.

\newpage
\begin{center}{References}
\begin{flushleft}
\begin{tabular}{r@{ }p{5.65in}}
$ [1] $ & E. M. Rastopchin, M. I. Svirin and G. N. Smirenkin, Yad. Fiz. 52 (1990) 1258\\
$[2]$ & S.T. Boneva, E.V. Vasilieva, Yu.P. Popov, A.M. Sukhovoj
 and V.A. Khitrov, Sov. J. Part. Nucl.  22(2) (1991) 232\\
 & S.T. Boneva et al., Sov. J. Part. Nucl.   22(6) (1991) 698\\
$[3]$ & A.M. Sukhovoj, V.A. Khitrov, Instrum. Exp. Tech. 27 (1984) 1071\\
$ [4] $ & V. A. Bondarenko, J. Honz\'atko, V. A. Khitrov, A. M. Sukhovoj
and I. Tomandl, Fizika B (Zagreb) 11 (2002) 201\\
$[5]$ & Yu.P. Popov, A.M. Sukhovoj, V.A. Khitrov and Yu.S. Yazvitsky,
Izv. AN SSSR, Ser. Fiz. 48 (1984) 1830\\
$[6]$ & J. Honz\'atko  et al., Nucl. Instr. and Meth.  A376 (1996) 434\\
$ [7]$ & E.V. Vasilieva, A.M.Sukhovoj and V.A. Khitrov,
Physics of Atomic Nuclei 64(2) (2001) 153\\
$ [8] $ & S.T. Boneva S.T., V.A. Khitrov,  A.M. Sukhovoj, Nucl.  Phys. A589 (1995) 293\\
$ [9] $ &  A.M. Sukhovoj, V.A. Khitrov, E.P. Grigoriev,
 Vopr. Atom. Nauki i Tech., Ser. Yd. Kons.
 1 (2001) 94\\
$[10]$ & S.G. Kadmenskij, V.P. Markushev and W.I. Furman,
Sov. J. Nucl. Phys.  37 (1983) 165\\
$[11]$ & W. Dilg, W. Schantl, H. Vonach and M. Uhl, Nucl. Phys. A217
(1973) 269\\
$ [12]$  & Ignatyuk A.V., Proc. IAEA Consultants meeting on the use of
nulear theory in neutron nuclear data evaluation (Trieste, Italy, December
1975): IAEA-190, 1 (1976) 211\\
$[13]$ & P. Axel, Phys. Rev. 126 (1962) 683\\
$ [14]$ &  J. Honzatko, V. A. Khitrov, C. Panteleev, A. M. Sukhovoj, I. Tomandl,
JINR Preprint E3-2003-88, Dubna, 2003, nucl-ex/0305020.\\
$ [15]$ &  A.M.Sukhovoj and V.A. Khitrov, Physics of Atomic Nuclei 62(1) (1999) 19\\

\end{tabular}
\end{flushleft}
 \end{center}

\newpage
\begin{center}
{\sl Table.
 Energies $E_1+E_2$ of cascades and their absolute intensities
$I_{\gamma\gamma}$ (\% per decay).
$E_f$ is the energy of the cascade final level}
\end{center}
\begin{center}
\begin{tabular}{|l|r|r|l|r|}  \hline
$E_1+E_2$, keV &  $E_f$, keV &
$I_{\gamma\gamma}^{exp}$ & $I_{\gamma\gamma}^{mod}$ & $I_{\gamma\gamma}^{best}$
\\ \hline
 9326.30  &0&    16.0(34)&  6.7 & 15.7\\
 8096.63  &1230& 15.3(11)&  7.2 & 15.5\\
 7568.00  &1758&   2.4(7)&  1.1 &  2.5\\
 7283.42  &2042&   3.3(16)& 2.4 &  5.5\\
 7269.39  &2057&   2.8(9)&  0.8 &  1.7\\
 7000.36  &2325+2328& 5.6(9)& 2.8& 5.4\\
 6923.08  &2403&   2.8(2)&  1.5 &  3.0\\
 6829.42  &2497&   [2]   &  0.5 &  0.7\\
 6648.95  &2677&   [1.5] & 1.0  &  2.0\\
 6588.29  &2738&   [4]   & 1.7  &  5.5\\\hline
sum       &    & 55.8(43)& 25.7 & 57.5\\
\hline
\end{tabular} \end{center}
{\it Note: $I_{\gamma\gamma}^{best}$ is the calculated mean value
for the ensembles
of random parameters $\rho$ and $k$
(its mean value and dispersion is shown in Fig.~3,4 by points with bars)
allowing reproduction [7]
of the cascade intensity distribution (Fig.~2) with experimental precision.
The mean-square scatter of each of these parameters for each final level of cascades equals 5 to 15\%.}

\begin{figure}
\begin{center}
\leavevmode
\epsfxsize=13.5cm
\epsfbox{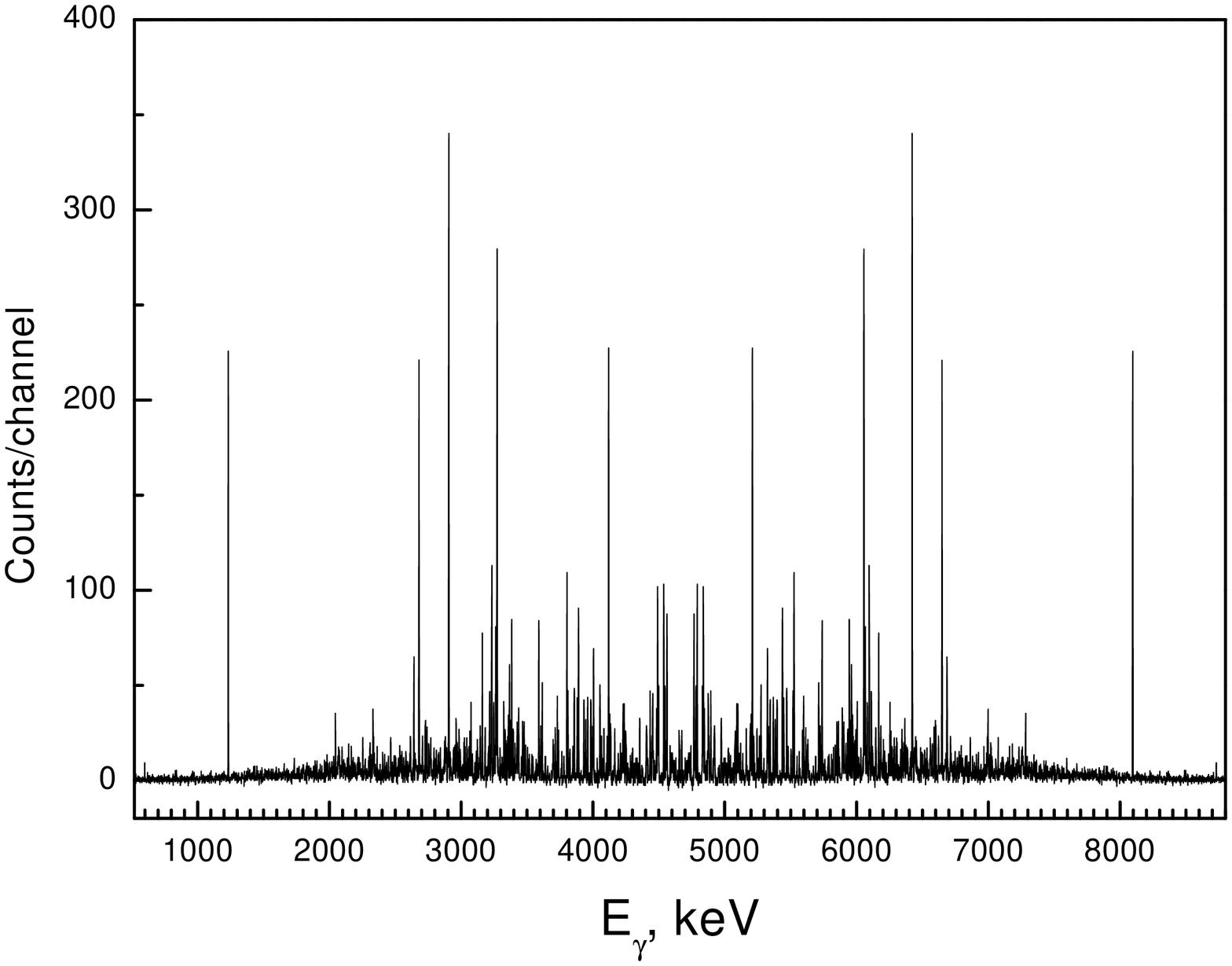}
\end{center}
\hspace{-0.8cm}

{\sl Fig.~1.~The intensity distribution of two-step cascades with
the total energy  $E_1+E_2$=9326 keV in $^{118}Sn$ (after
background subtraction and correction for efficiency of
registration of cascades)}.
\end{figure}
\newpage
\begin{figure}
\begin{center}
\leavevmode
\epsfxsize=13.5cm
\epsfbox{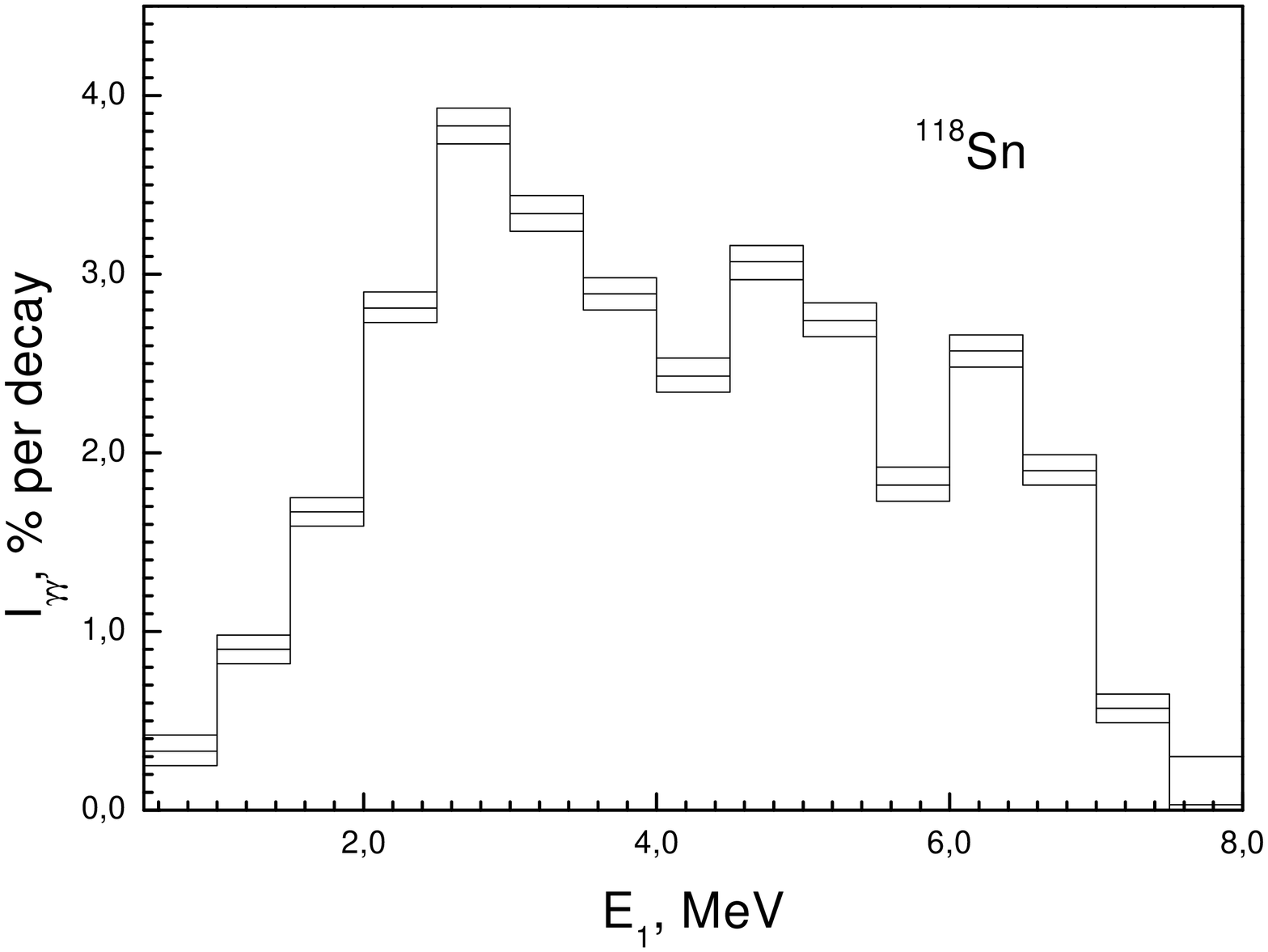}
\end{center}
\hspace{-0.8cm}

{\sl Fig.~2.~Experimental distribution of the total intensity of
the two-step cascades terminating at the ground and first excited
states of $^{118}Sn$ in function of energy of their primary
transitions $E_1$ The ordinary statistical error is shown.}
\end{figure}
\newpage
\begin{figure}
\begin{center}
\leavevmode
\epsfxsize=11cm
\epsfbox{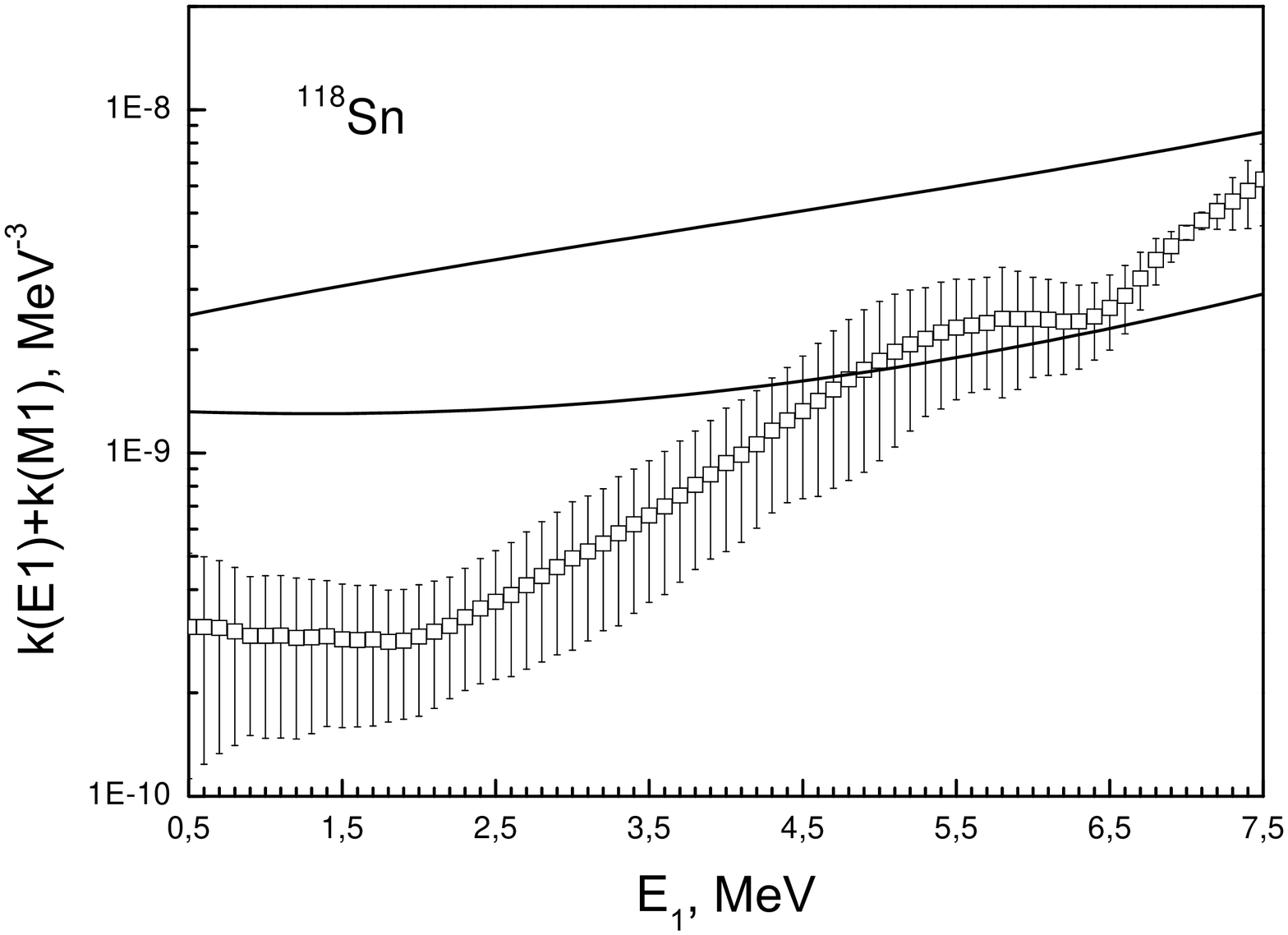}
\end{center}
\hspace{-0.8cm}

{\sl Fig.~3.~The sum of the probable radiative strength functions of
$E1$ and $M1$ transitions (with estimated errors).
 The upper and lower solid curves represent predictions of the
models [13] and [10], respectively (the value $k(M1)=const$ normalized
to the experiment at $E_{\gamma} \approx B_n$ is added).}
\end{figure}
\newpage
\begin{figure}[h]
\begin{center}
\leavevmode
\epsfxsize=11cm
\epsfbox{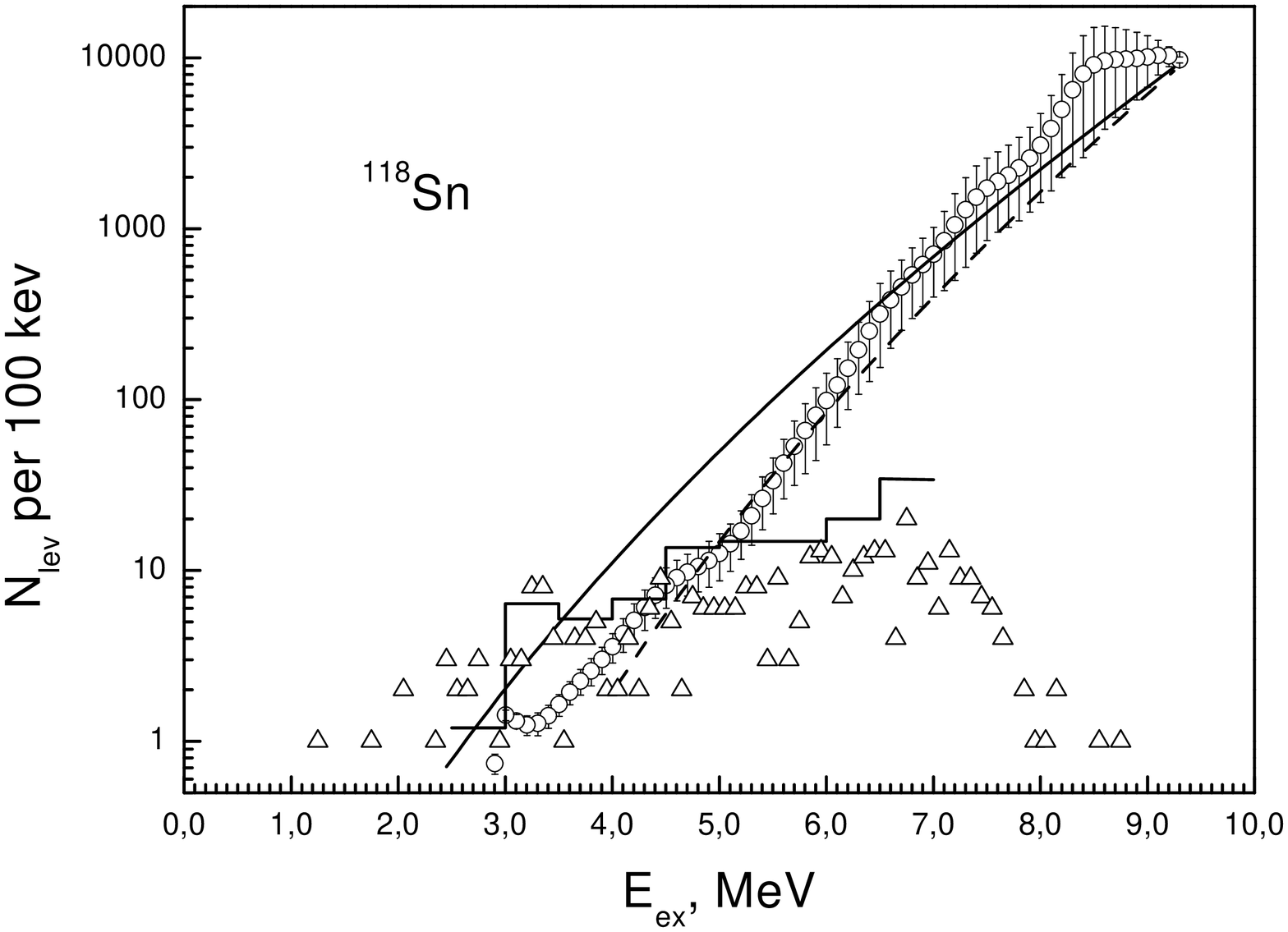}
\end{center}
\hspace{-0.8cm}

{\sl Fig. 4.~
The number of levels for $0^{\pm} \leq J^{\pi} \leq 2^{\pm}$
 with their dispersion (circles with bars). The histogram represents the
data of analysis [15], triangles show the observed in experiment
number of intermediate levels of intense cascades. Solid and dashed lines
represent predictions of the models [11] and [12], respectively.}
\end{figure}

\end{document}